\documentclass[final,5p,times,twocolumn]{elsarticle}

\usepackage{amssymb}
\usepackage{amsmath}

\usepackage{lineno} 
\usepackage{hyperref}

\journal{Nuclear Physics B}

\begin{document}

\begin{frontmatter}




\title{Cryogenic performance of Silicon Photo-Multipliers at LN$_2$ temperature and at 3 K in the RES-NOVA cryostat}



\author[unipv,infnpv]{R. Rossini}
\author[infnpv]{M. Rossella}
\author[infnmb]{M. Clemenza}
\author[infnpv]{S. Copello}
\author[infnpv]{C. De Vecchi}
\author[infnmb]{S. Dell'Oro}
\author[unimb,infnmb]{L. Gironi}
\author[infnmb]{C. Gotti}
\author[unipv,infnpv]{L. Malliolio}
\author[unipv,infnpv]{N. Manenti}
\author[unipv,infnpv]{A. Menegolli}
\author[unimb,infnmb]{L. Pattavina}
\author[unimb,infnmb]{D. Peracchi}
\author[unimb,infnmb]{S. Quitadamo}
\author[infnpv]{M.C. Prata}
\author[infnpv]{G.L. Raselli}
\author[unimb,infnmb]{L. Trombetta}

\affiliation[unipv]{
    organization={Department of Physics "A. Volta", University of Pavia},
    city={Pavia},
    country={Italy}
}

\affiliation[infnpv]{
    organization={INFN sezione di Pavia},
    city={Pavia},
    country={Italy}
}

\affiliation[infnmb]{
    organization={INFN sezione di Milano-Bicocca},
    city={Milan},
    country={Italy}
}

\affiliation[unimb]{
    organization={Department of Physics "G. Occhialini", University of Milano-Bicocca},
    city={Milan},
    country={Italy}
}

\begin{abstract}
Silicon Photo-Multipliers (SiPMs) are semiconductor-based photodetectors that offer performance comparable to that of traditional Photo-Multiplier Tubes (PMTs), while providing several advantages, including low operating voltage, immunity to magnetic fields, compact size, and mechanical robustness. These features have led to their widespread adoption in scintillator-based detectors for both fundamental and applied physics. In recent years, there has been a growing interest in the detection of scintillation light at cryogenic temperatures in low-temperature experiments based on bolometric detectors, such as RES-NOVA and CUPID, as well as in emerging quantum applications. In the RES-NOVA experiment, radiopure PbWO$_4$ crystals are used to detect supernova neutrinos. The scintillation of PbWO$_4$ can only be evaluated at cryogenic temperatures, as the scintillation light yield is negligible at room temperature. We present measurements of the performance at cryogenic temperatures of Hamamatsu S13360-1350CS SiPMs in terms of breakdown voltage and gain, at 77~K (in a liquid nitrogen bath) and at 3~K. The cooling down to 3~K has been carried out using a Proteox LX cryostat from Oxford Instruments. 
\end{abstract}




\end{frontmatter}



\section{Introduction}

The detection of neutrinos from core-collapse Supernovae represents a powerful tool for investigating the mechanisms governing stellar evolution and gravitational collapse. RES-NOVA is an experiment aiming to detect astrophysical neutrinos, particularly those emitted by Supernovae, through Coherent Elastic Neutrino-Nucleus Scattering (CE$\nu$NS) \cite{resnova2020}. The proposed detector consists of an array of cryogenic calorimeters based on lead tungstate (PbWO$_4$) crystals produced from archaeological lead. The high neutrino interaction cross-section of lead, combined with the exceptional radiopurity of the archaeological material, provides the potential for high sensitivity to low-energy nuclear recoils in an ultra-low-background environment \cite{resnova2022,resnova2023}.

The RES-NOVA detection technique relies on operating PbWO$_4$ crystals at cryogenic temperatures, at the 10 mK scale, where the energy deposited by particle interactions can be measured through the resulting phonon signals \cite{resnova2026}. In addition to the calorimetric channel, the scintillation light emitted by the crystals provides a complementary source of information for particle identification and background discrimination. The simultaneous detection of heat and light can help distinguish nuclear recoils from electron recoils, thereby improving the rejection of background events. The development of a suitable photosensor to the test the scintillation properties of PbWO$_4$ at low temperatures is therefore an important aspect of the detector instrumentation.

In particular, PbWO$_4$ crystals exhibit negligible scintillation at room temperature, whereas their light yield increases as the temperature decreases, reaching a sort of plateau at liquid nitrogen temperature (77 K)~\cite{pbwo4}. This study goes in the direction of setting up a facility for a quick optical characterisation of the crystals at temperatures below 77 K.

Silicon Photo-Multipliers (SiPMs) are promising candidates for this application due to their high photon detection efficiency, compact dimensions, low operating voltage and capability of detecting weak light signals, down to the single photon level. However, their performance at cryogenic temperatures must be thoroughly investigated before their implementation in a detector operating at temperatures well below 1 K~\cite{cervi1,cervi2}. In particular, temperature-dependent variations in the breakdown voltage, gain, dark count rate and correlated noise can affect the stability and reliability of the optical readout. The study of these parameters is essential to assess the feasibility of using SiPMs for the characterization of the RES-NOVA crystals.

\begin{figure}
    \centering
    \includegraphics[width=\linewidth]{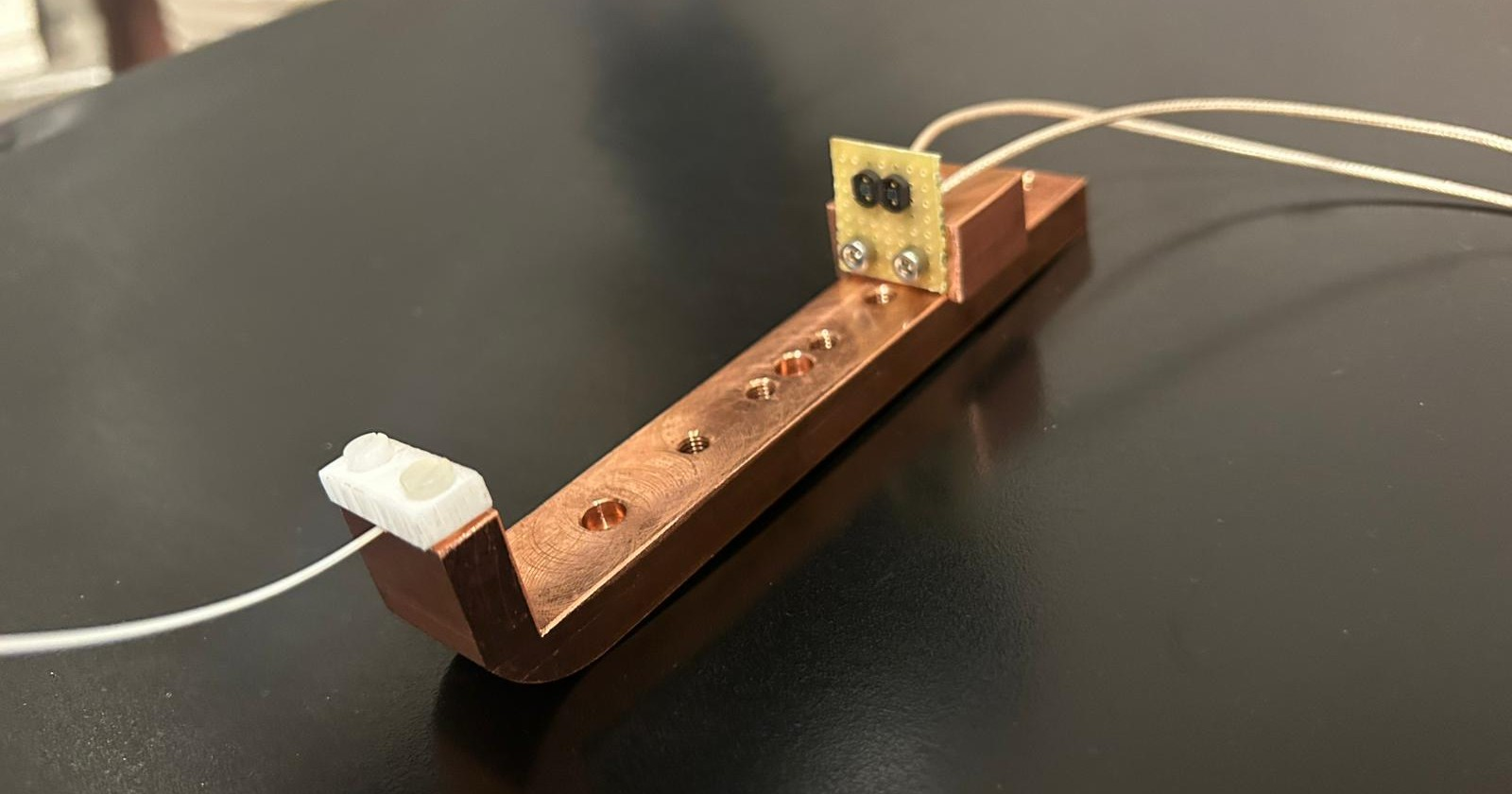}
    \caption{Setup for the SiPM characterisation in the cryostat: two SiPMs are positioned at a variable distance (up to 10 cm) from an optical fibre used to feed light to the devices. Copper is used for the support to optimise thermalisation.}
    \label{fig:sipm}
\end{figure}

In this work, the low-temperature performance of selected SiPM models is investigated with the aim of evaluating their suitability as photosensors for PbWO$_4$ scintillating crystals. The devices are characterized under controlled thermal conditions, with particular attention to the evolution of their electrical and optical properties as a function of temperature and applied bias voltage. The measurements focus on the parameters relevant to low-light detection, such as the breakdown voltage and the gain, and on identifying possible limitations associated with cryogenic operation. This study 
contributes to the development of photosensor solutions also for cryogenic scintillating detectors.

\section{The Oxford Instrument PT cryostat at the University of Milano-Bicocca}

The cryogenic characterization of the SiPMs was performed at the University of Milano-Bicocca using an Oxford Instruments cryogen-free dilution refrigerator, equipped with two pulse-tube (PT) precooling systems. The cryostat is designed to provide stable operating conditions at millikelvin temperatures and is used for the experimental activities of the RES-NOVA project. The cooling system combines a closed-cycle pulse-tube refrigerator, which provides the initial cooling of the cryostat, with a $^{3}$He/$^{4}$He dilution circuit to reach ultra-low temperatures. The cryostat features several thermal stages, including intermediate temperature stages at approximately 40~K and 3~K, a still stage at around 700~mK, a cold plate at approximately 100~mK, down to the mixing chamber, where temperatures of about 10~mK can be achieved. Its cooling power exceeds 450~$\mu$W at 100~mK, allowing the operation of experimental setups requiring high thermal stability at cryogenic temperatures. The system provides a suitable environment for investigating the temperature-dependent electrical and optical properties of SiPMs and assessing their potential use as photosensors for the characterization of the scintillation light yield of PbWO$_4$ crystals of RES-NOVA, and possibly of other experiment employing scintillating low-temperature calorimeters such as CUPID \cite{cupid2025}.

\section{Experimental setup}

\begin{figure}[ht]
    \centering
    \includegraphics[width=\linewidth]{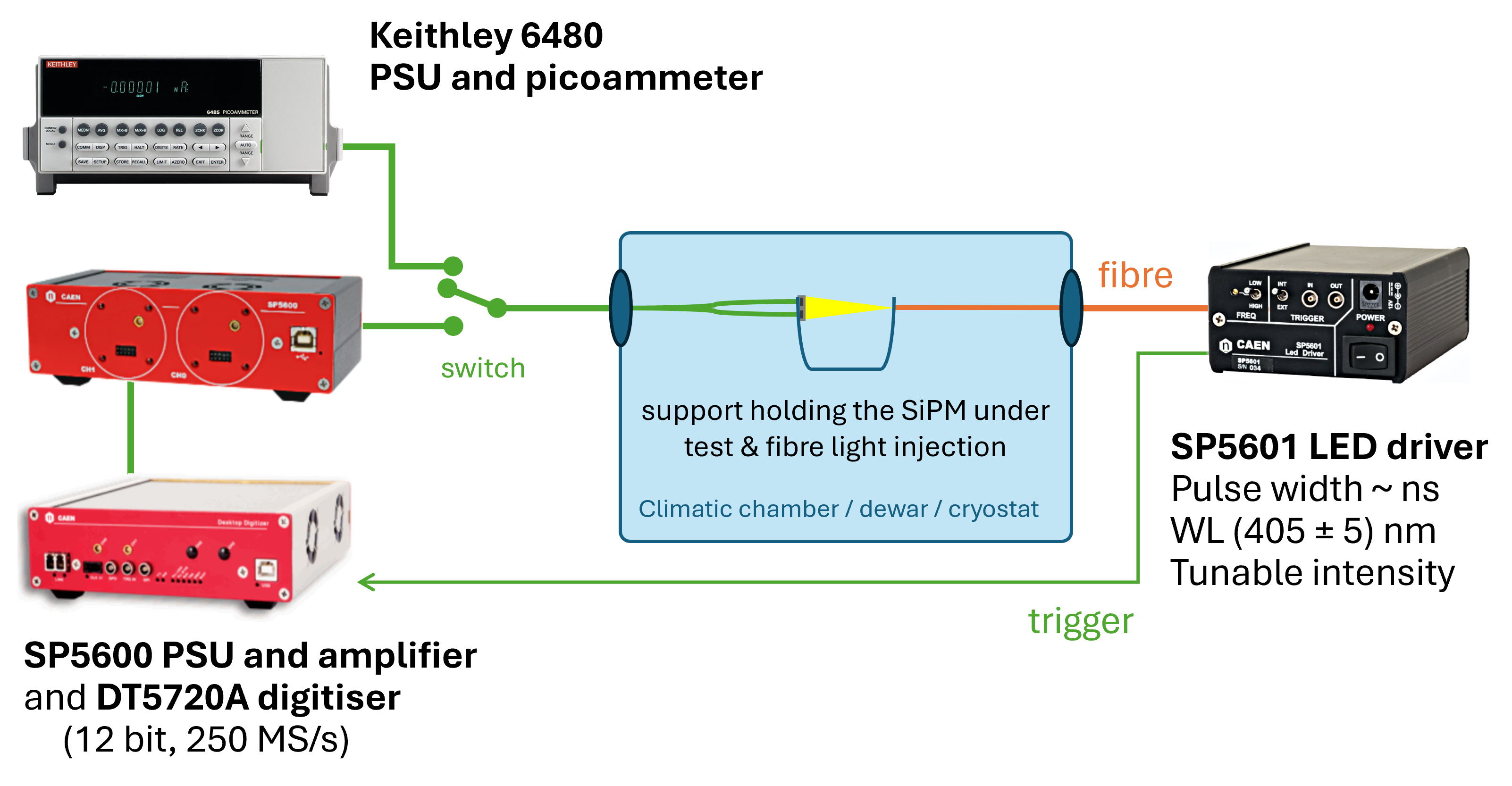}
    \caption{Sketch of the setup adopted for the measurement inside the climatic chamber.}
    \label{fig:climatic}
\end{figure}

\begin{figure}
    \centering
    \includegraphics[width=\linewidth]{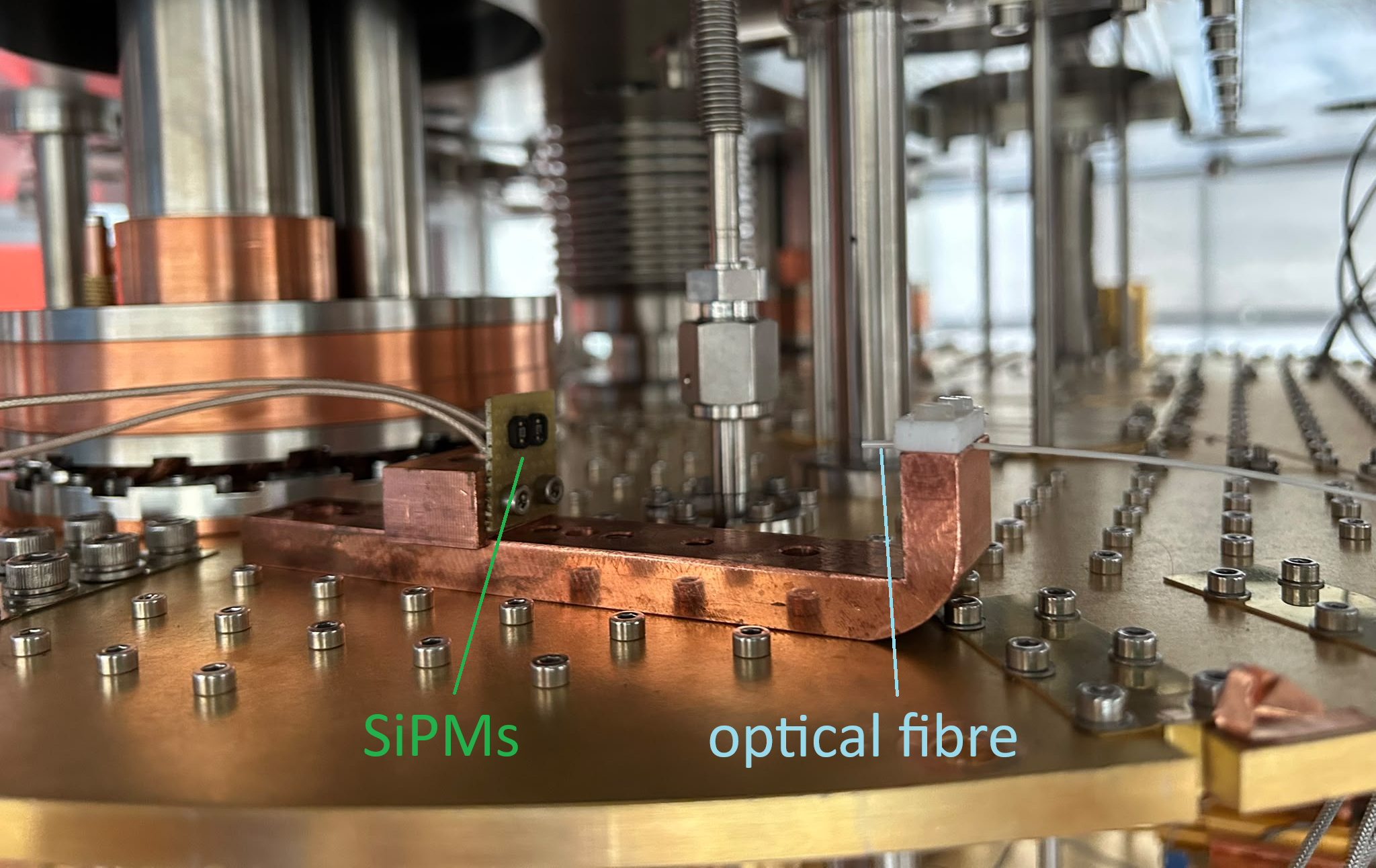}
    \caption{Experimental setup mounted in the cryostat: the two Hamamatsu S13360-1350 SiPMs under test are mounted on a board and connected to the electronics through two coaxial cables, whereas the optical fibre is clamped in front of them.}
    \label{fig:setup}
\end{figure}

\begin{figure*}
    \centering
    \includegraphics[height=0.34\linewidth]{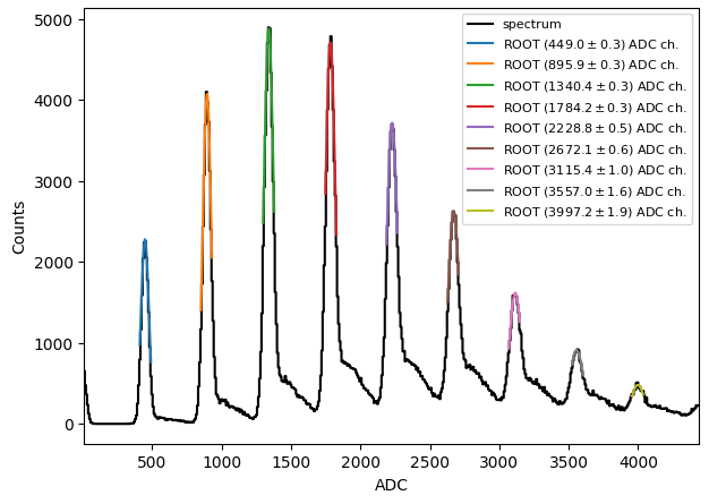}
    \includegraphics[height=0.38\linewidth]{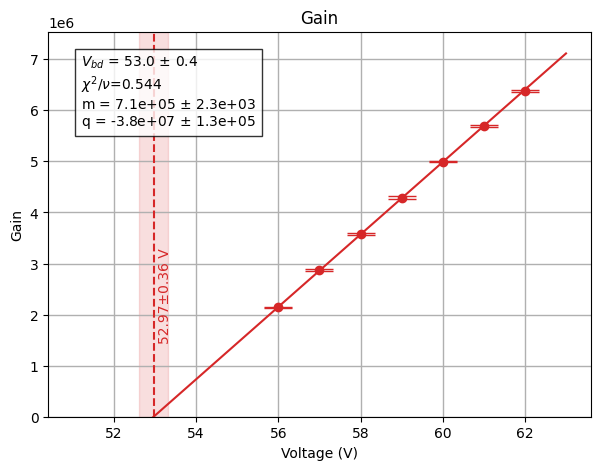}
    \caption{Spectrum of SiPM (left), expressed in ADC counts, showing the discrete structure of the photo-sensor, and SiPM gain (right) evaluated as a function of the biasing voltage, fitted with a linear function. The voltage value corresponding to zero gain in the fit is the breakdown voltage.}
    \label{fig:sipm_spectrum_gain}
\end{figure*}

The setup consists of two Hamamatsu S13360-1350 SiPMs, featuring an active area of $1.3 \times 1.3~\mathrm{mm}^2$ and a microcell pitch of $50~\mu\mathrm{m}$, mounted on a copper support designed and manufactured at the INFN Pavia mechanical workshop. The support allows seven different SiPM-to-fibre distances and features three mounting holes for installation inside the cryostat. The optical excitation is provided by a fibre with a $50~\mu\mathrm{m}$ core and a ceramic ferrule, while the SiPMs are connected to the external readout electronics through RG178 coaxial cables, see Fig.~\ref{fig:sipm}. The optical fibre and electrical signals are routed through dedicated cryostat feedthroughs. An FC/PC feedthrough is used for the optical fibre, while the electrical connections are provided by a four-pin feedthrough, accommodating the two coaxial cable cores and their corresponding braids.

The external instrumentation is based on a modified CAEN SP5600E Educational Photon Kit. The readout system comprises a CAEN PSAU SP5600 power supply and amplification unit, and a DT5720A digitizer, used for the evaluation of the SiPM noise as a function of the discriminator threshold, and the measurement of the gain through detection of single photons. A CAEN SP5601 LED driver provides the optical excitation required for gain measurements, see Fig.~\ref{fig:climatic}. The resulting single-photon spectra are used to determine the SiPM gain and extract the breakdown voltage ($V_{\mathrm{bd}}$). In addition, the current--voltage (I--V) characteristics are measured using a Keithley 6487 picoammeter, allowing an independent determination of the breakdown voltage. The combination of these measurement techniques provides complementary information on the electrical and optical performance of the devices.

Preliminary measurements were initially performed at the Department of Physics of the University of Pavia using a climatic chamber (F.lli Galli model Genviro-030LC with an allowed temperature range from $-70^{\circ}\mathrm{C}$ to $+180^{\circ}\mathrm{C}$) over a temperature range from $-50^{\circ}\mathrm{C}$ to $+25^{\circ}\mathrm{C}$. For the measurements in the climatic chamber, a more compact setup was deployed, composed of a single SiPM and the optical fibre right in front of it, for optimised SiPM illumination.

These tests included dark rate measurements and optical excitation with an LED. Further measurements were carried out at $77~\mathrm{K}$ using a liquid nitrogen bath, where the I--V characteristics were investigated. Optical excitation was not tested in this configuration because the setup was immersed in liquid nitrogen. Following these preliminary tests, the assembly was installed in the cryostat at the University of Milano-Bicocca, with the setup reaching an operating temperature of $3~\mathrm{K}$ during the installation and initial testing campaign, see Fig.~\ref{fig:setup}. This configuration enables the investigation of the SiPM response under cryogenic conditions, providing the basis for subsequent measurements at low temperatures.

\section{First test in a climatic chamber}

\begin{figure}
    \centering
    \includegraphics[width=\linewidth]{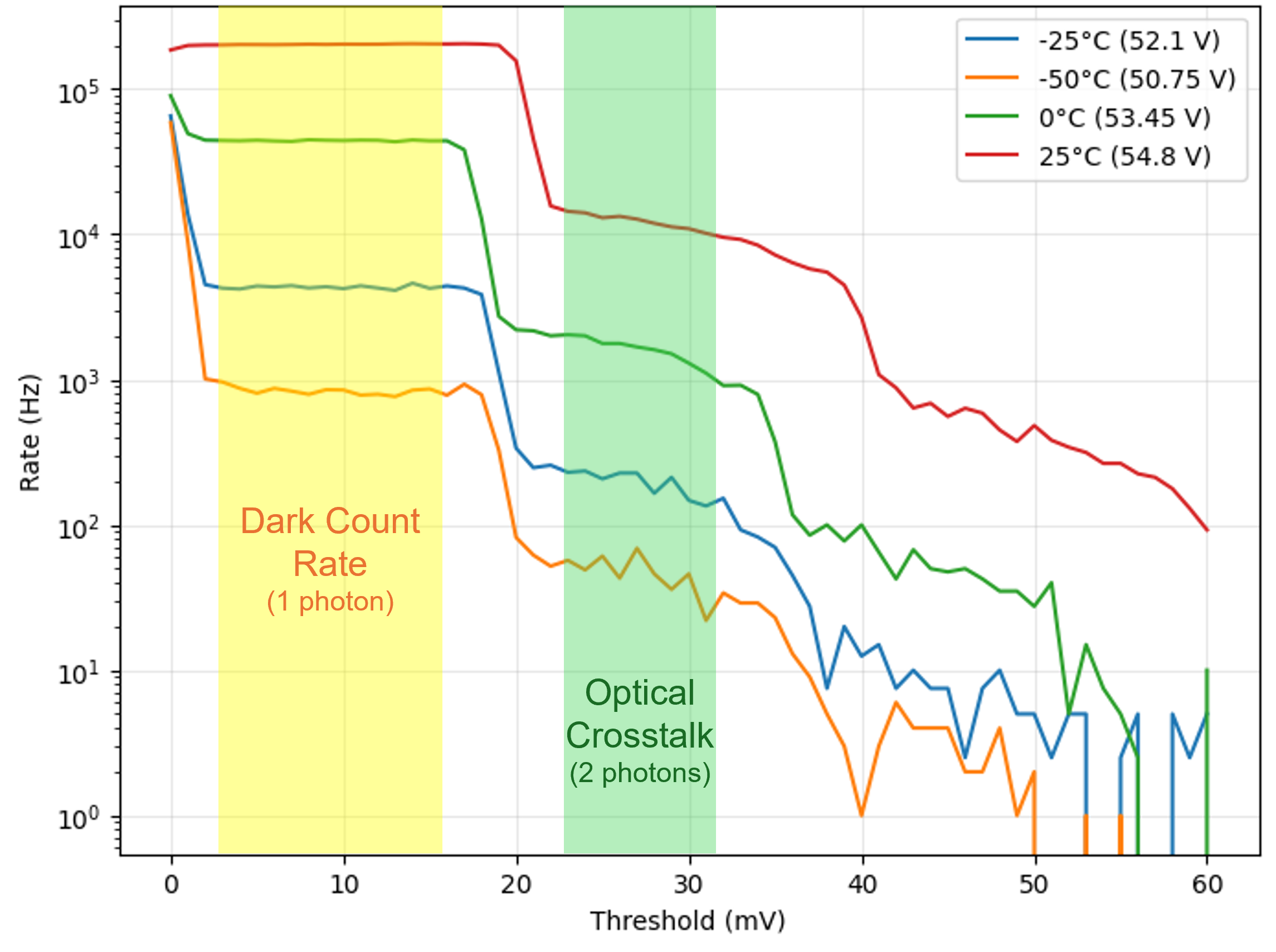}
    \caption{Staircase plot as a function of the temperature in the climatic chamber. The SiPM is biased at the same overvoltage and let without illumination. The rate of signals is measured as a function of the trigger threshold. The first plateau corresponds to the dark count rate (triggering on single photon signals), the second one to the optical crosstalk rate (triggering on double photons). Both parameters visibly decrease with the temperature.}
    \label{fig:dcr_thr}
\end{figure}

\begin{figure}
    \centering
    \includegraphics[width=.7\linewidth]{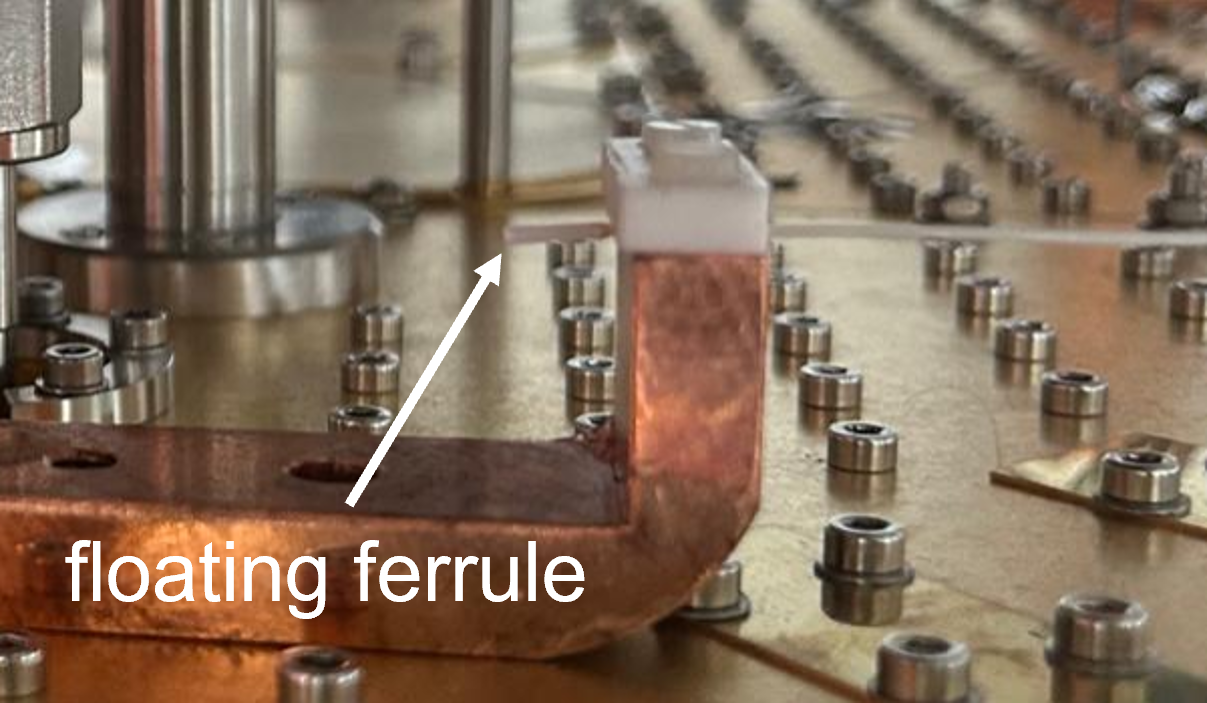}
    \caption{Detail of the fibre clamping, with the ferrule left floating to avoid mechanical stress. }
    \label{fig:fibre}
\end{figure}

\begin{figure*}
    \centering
    \includegraphics[height=0.33\linewidth]{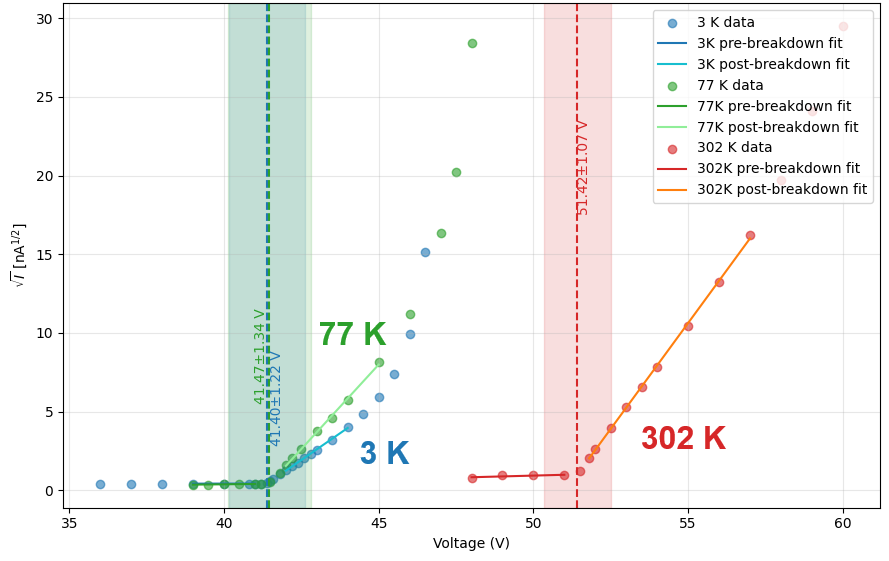}
    \includegraphics[height=0.33\linewidth]{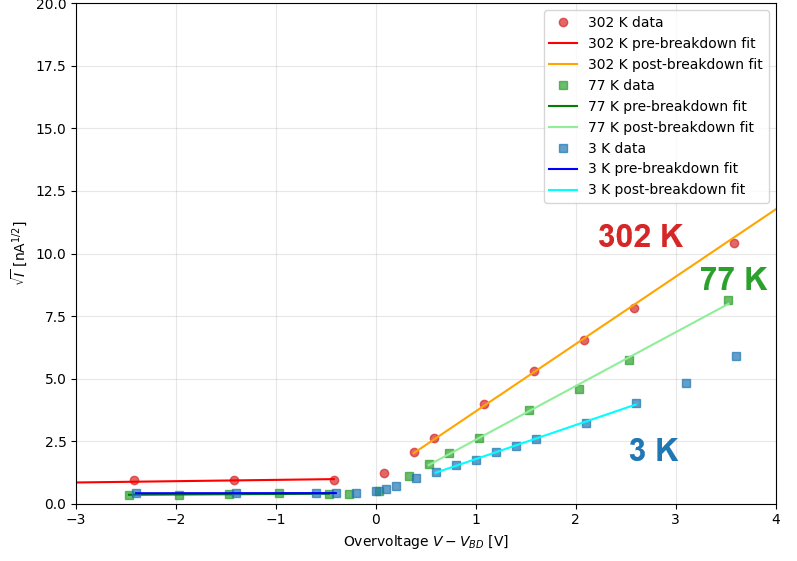}
    \caption{$\sqrt{I}$--V characteristic curve (left) fitted with two linear functions before and after the breakdown value, which is obtained from the intersection of the two lines, at 302 K, 77 K, and 3 K. In the right panel, the same curves are translated to show the overvoltage ($V \to V-V_{bd}$), showing the expected decrease in dark current as temperature decreases.}
    \label{fig:cryoIV}
\end{figure*}

Preliminary measurements inside the climatic chamber have been performed at four different temperatures: $T = -50~^\circ$C, $T = -25~^\circ$C, $T = 0~^\circ$C, and $T = +25~^\circ$C for one of the two SiPMs mounted in the setup.
The SiPM gain is evaluated from the sensor's photon spectrum, which is obtained from the charge pulses recorded by the digitizer and integrated over a fixed 256 ns time window. After converting the measured values using a calibration factor, $\mathrm{Cal} = 39~\mathrm{fC/ADC} \cdot 1/G_A$, which accounts for the digitizer specifications and the gain amplification $G_A$ of the SP5600, the SiPM gain is calculated by measuring the distance between adjacent peaks in the photon spectrum (in ADC channels), according to the formula: $\mathrm{Gain} = (\mathrm{ADC~ch} \cdot \mathrm{Cal})/e$, see the left panel in Fig.~\ref{fig:sipm_spectrum_gain}.

As an example, the right panel in Fig.~\ref{fig:sipm_spectrum_gain} shows the gain as a function of the bias voltage for the Hamamatsu S13360-1350 SiPM at room temperature. The breakdown voltage is extrapolated by fitting the gain as a function of the bias voltage and determining the voltage at which the gain becomes zero.

Finally, the Dark Count Rate (DCR) is evaluated by setting the detection threshold to 0.5 photoelectrons (p.e.) and counting the number of pulses exceeding this threshold. This provides a measure of the rate of dark-count events, i.e., spurious detections occurring in the absence of incident photons. Fig.~\ref{fig:dcr_thr} shows the so-called staircase plot, which represents the dark count rate as a function of the applied threshold at the four different temperatures, all measured at the same overvoltage. As expected, the dark count rate decreases with decreasing temperature.

\section{Cryogenic tests}

After the first validation of the setup adopted for the cryogenic measurements, a first test was carried out, again in Pavia laboratory, at $77~\mathrm{K}$ using a liquid nitrogen bath, to test the mechanical behaviour of the system and to study I--V characteristics. To prevent mechanical stress on the optical fibre, its end was clamped, while the ferrule was left floating. Fig.~\ref{fig:fibre} shows the detail of the fibre clamping in the final setup installed in the dilution cryostat (Fig.\ref{fig:setup}). As expected, the breakdown voltage decreases at cryogenic temperature, being measured at $\sim$ 41.5 V at $77~\mathrm{K}$, versus the $\sim$ 51.5 V measured at room temperature (see Fig.~\ref{fig:cryoIV}): in fact, at higher temperatures, increased scattering reduces the energy gained by charge carriers, requiring a stronger electric field to initiate an avalanche. 
Conversely, at lower temperatures, charge carriers experience minor energy-loss events due to scattering with the crystal lattice, allowing them to gain more energy from the electric field between successive scattering events. As a result, the probability of impact ionization increases, and a lower electric field is required to trigger avalanche multiplication. 

The final test of this first approach to evaluating the cryogenic performance of the SiPM models selected was carried out in the cryostat at the University of Milano-Bicocca, where the setup reached an operating temperature of $3~\mathrm{K}$. The I--V curve, shown in Fig.~\ref{fig:cryoIV}, allowed the breakdown voltage to be estimated at approximately $41.5~\mathrm{V}$ for both SiPMs mounted in the setup.

The observation that the breakdown voltage remains unchanged between $77~\mathrm{K}$ and $3~\mathrm{K}$ is consistent with previous cryogenic measurements reported by other researchers~\cite{eth_cryo} using Hamamatsu S13370 (VUV4) SiPMs. In that study, the authors observed a linear decrease in $V_{bd}$ with decreasing temperature between room temperature and $150~\mathrm{K}$. At lower temperatures, however, the breakdown voltage deviates from this linear trend and approaches a plateau slightly below $42~\mathrm{V}$. This behaviour can be explained by the temperature dependence of avalanche formation in silicon~\cite{cryosipm}.

Our measurements are therefore in good agreement with previous observations and provide a validation of the experimental setup adopted for these tests. New cryogenic measurement campaigns are consequently planned, taking advantage of the possibility of moving the setup to an intermediate stage of the dilution cryostat, allowing for a characterization of the SiPMs at $38~\mathrm{K}$. Preliminary measurements had already been performed by mounting the setup on this stage. However, uncertainties regarding the thermalisation of the setup with the cryogenic system raised doubts as to whether the setup had actually reached a temperature of $38~\mathrm{K}$. A more accurate measurement at $38~\mathrm{K}$ will be then carried out during the next experimental campaign using the dilution cryostat.

\section{Conclusion}
This preliminary characterisation goes in the direction of setting up a facility for a quick optical characterisation of the RES-NOVA PbWO$_4$ crystals, for which scintillation is strongly suppressed at room temperature, and is much more prominent below 0 $^\circ$C \cite{pbwo4}. 

The test setup is currently operational and has been tested both at liquid nitrogen temperature and at 3 K in the cryostat, yielding promising results. Further cryogenic measurement campaigns are already planned to investigate the performance of the SiPMs at further cryogenic temperatures.

\bibliographystyle{elsarticle-num}

\end{document}